\documentclass[useAMS,usenatbib]{mn2e}

\def\arcsec{$^{\prime}$$^{\prime}$}

\newcommand{\isot}[1]{$^{#1}$}
\newcommand{\hii}{H{\sc ii}}

\newcommand{\rev}{}

\usepackage[T1]{fontenc}
\usepackage{aecompl}

\newcommand{\aj}{{AJ}}                   
\newcommand{\araa}{{ARA\&A}}             
\newcommand{\apj}{{ApJ}}                 
\newcommand{\apjl}{{ApJ}}                
\newcommand{\apjs}{{ApJS}}               
\newcommand{\aap}{{A\&A}}                
\newcommand{\pasp}{{PASP}}
\newcommand{\pasj}{{PASJ}}
\newcommand{\mnras}{{MNRAS}}
\newcommand{\actaa}{{Acta Astron.}}

\usepackage[usenames,dvipsnames]{xcolor}
\usepackage{graphicx}
\usepackage{color}
\usepackage{txfonts}
\newcommand{\colines}{$^{12}$CO, $^{13}$CO and C$^{18}$O (J=1$-$0)}
\title[Pillar Dynamics in Vulpecula]{Cloud Disruption via Ionized Feedback: Tracing Pillar Dynamics in Vulpecula}

\author[Klaassen et al.]{P.D. Klaassen$^{1}$\thanks{e-mail: klaassen@strw.leidenuniv.nl}, 	J.C. Mottram$^{1}$, J.E. Dale$^{2}$ and A. Juhasz$^{1}$\\
$^{1}$Leiden Observatory, Leiden University, PO Box 9513, 2300 RA Leiden, The Netherlands\\
$^{2}$Excellence Cluster `Universe', Boltzmannstr. 2, 85748 Garching, Germany
}

\begin{document}

   \date{}
   
   \pagerange{\pageref{firstpage}--\pageref{lastpage}} \pubyear{2014}

\maketitle

\label{firstpage}

\begin{abstract}
The major physical processes responsible for shaping and sculpting pillars in the clouds surrounding massive stars (i.e. the `Pillars of Creation') are now being robustly incorporated into models quantifying the ionizing radiation from massive stars. The detailed gas dynamics within these pillars can now be compared with observations. Our goal is to quantify the gas dynamics in a pillar being sculpted by a nearby massive star. To do this, we use the CO, \isot{13}CO, and C\isot{18}O J=1-0 emission towards a pillar in the Vulpecula Rift. These data are a combination of CARMA and FCRAO observations providing high resolution ($\sim5ее$) imaging of large scale pillar structures ($>100''$).  We find that this cold ($\sim18$ K), low density material ($8\times10^3$ cm$^{-3}$) material is fragmenting on Jeans scales, has very low velocity dispersions ($\sim0.5$ km s$^{-1}$), and appears to be moving away from the ionizing source. We are able to draw direct comparisons with three models from the literature, and find that those with lower velocity dispersions best fit our data, although the dynamics of any one model do not completely agree with our observations. We do however, find that our observed pillar exhibits many of the characteristics expected from simulations.    
\end{abstract}


%

\section{Introduction}

 The appearance of many young star-forming regions is dominated by bubble structures of sizes ranging from a few tenths of a parsec to many tens of parsecs \citep{Churchwell06,Simpson12}. These structures are generated by the ionizing radiation and winds (and sometimes probably also the supernovae) of massive stars. Common but by no means universal accompaniments to bubbles are pillar-like or ``elephants'  trunk'' structures projecting inwards from the bubble walls toward the OB-stars. These were first discovered as reflection nebulae in the optical \citep{Minkowski1949}, with the most well-known being the so-called ``Pillars of Creation'' \citep{Hester1996}. Over time, observations at longer wavelengths have confirmed that these structures are cold, dense and molecular \citep[e.g.][]{Pound1998}. However, it is only in the past few years that high-resolution multi-wavelength surveys have allowed comprehensive study of both these pillars and the star formation going on inside and around them \citep[e.g.][]{Billot2010,Chauhan2011,Preibisch2012}. Similarly, while their shape and their relation to nearby O-stars lead \citet{Frieman1954} to suggest instability as important in their formation, recent advances in numerical modelling now allow two- and three-dimensional models of pillars and of whole bubbles as well \citep[e.g.][]{Gritschneder10,Haworth2012,Tremblin12b,Dale13a}.

  Whether photoionization models contain perturbations in the ionizing radiation field \citep[e.g][]{Garcia96}, density structures in the ambient material \citep[e.g.][]{Tremblin12a,Williams01}, or turbulence within the gas \citep[e.g.][]{Gritschneder10}, or any combination thereof, they tend to form realistic pillar like structures.  \citet{Haworth2012} even show that for thin shell regions, only a relatively small ionizing flux is required to produce pillars.

 \cite{Walch12,Walch13}  modelled the influence of central ionizing radiation sources on clouds with various initial fractal dimensions. They found that the efficiency of pillar formation was a function of the fractal dimension. Low fractal dimensions resulted in extended, relatively smooth shell-like structures, whereas higher fractal dimensions produced large numbers of pillar structures from the shadowing effects of small dense clumps.
  
  \cite{Dale12a,Dale12b,Dale13a} and \cite{Dale13b} modelled the propagation of ionizing radiation from O--stars from {\it inside} turbulent molecular clouds to examine the impact of feedback on the cloud structure and dynamics and on the ongoing star formation process within the clouds. The simulated clouds have a filamentary structure generated by the interaction of turbulent flows, and gas flows along the filaments, so they are also accretion flows. The most massive clusters and stars therefore tend to form at filament junctions. Ionization feedback in these simulations often produces prominent bubbles on scales up to a few $\times10$pc. In several of \citet{Dale12b} calculations, most conspicuously in their Run I, pronounced pillar structures emerge naturally from the erosion of the filaments/accretion flows by the massive stars. These pillars tend to have conical shapes.

That pillar morphologies are generally being reproduced by all of these models suggests that the primary physical processes and initial conditions required to form pillars are being included in simulations.  But how do the gas kinematics compare between observations and model predictions?  \citet{Tremblin13} make comparisons between large scale molecular gas motions in the Rosette and Eagle nebulae  and their pillar creation models. They find that the gas motions were also in good agreement with their modelled pillars at early and late stages of development.  However, their observed spectra were averaged over a pillar, and do not address motions within the pillars themselves.  To do this requires resolved observations of the gas motions within pillars.

To test these models of how pillars are formed in the presence of a nearby massive star, we have observed the gas dynamics in such a pillar in the Vulpecula rift.

	The Vulpecula Rift molecular cloud complex is a large, relatively evolved star forming region located in the Sagittarius spiral arm at a distance of 2.3\,kpc \citep{Massey1995}. It is dominated by three evolved \hii\, regions, Sharpless 86, 87 and 88 \citep{Sharpless1959} as well as the OB association Vul\,OB1 \citep{Massey1995,Reed2003,Bica2008}. This consists of $\sim$90 OB stars, approximately one third of which are in the open cluster NGC\,6823 which has an age of 3$\pm$1\,Myr \citep{Pigulski2000}. These ionizing sources have a considerable effect on the structure of the molecular material and star formation in the region. \citet{Billot2011} found that the degree of clustering of sources in \textit{Herschel} Hi-GAL survey \citep{Molinari2010} observations around the \hii\, regions decreases with increasing wavelength and \citet{Billot2010} identify 14 pillars pointing at either NGC\,6823 or other Vul\,OB1 members. 
	
	There are many protostars in the region, some driving very powerful outflows \cite[e.g.][]{Beuther2002} including one associated with a small cluster of MYSOs at the base of a pillar \citep{Mottram2012}. There is no sign of an age gradient in the YSO population across the region \citep{Billot2010}, so star formation has not yet been completely quenched. This combination of factors; main-sequence OB stars, evolved \hii\, regions, significant remaining molecular gas and dust, and ongoing steady star formation, all make this region ideal for studying feedback and the properties of pillars.

Here we present high resolution ($\sim 5$\arcsec) observations of the large scale CO emission towards a pillar in the Vulpecula rift \citep[Vulp3 using the terminology of][]{Billot2010} 16 pc from the ionizing source.  Below, in Section \ref{sec:obs} we present our CARMA and FCRAO observations and how they were combined, and in Section \ref{sec:results} we present our observational results. In Section \ref{sec:discussion} we discuss the implications of these results in terms of the physical processes at work around and within the pillar, as well as comparing our results to the quantifiable predictions from some of the models described above.  In Section \ref{sec:conclusions} we conclude.

\section{Observations}
\label{sec:obs}

The data presented here comprise CARMA and FCRAO observations of \colines\, towards a pillar in the Vulpecula Rift (R.A = 19:44:59, DEC = 23:47:14.6). These observations were  jointly inverted into the image plane to recover the large scale structures at high angular resolution.  Here we describe the data quality and properties as well as how the datasets were reduced independently for the CARMA (Section \ref{sec:carma}) and FCRAO (Section \ref{sec:FCRAO}) observations. We then describe how the datasets were combined in Section \ref{sec:datacombo}.

\subsection{CARMA data}
\label{sec:carma}

The CARMA data were observed in the D configuration in seven observing blocks between March 30 and June 22 2013, for project c1072.  The calibrators used for each observing block are listed in Table \ref{tab:calibrators}.  The 19 point mosaic was observed with the J=1-0 transition of CO (115.27 GHz), $^{13}$CO (110.20 GHz) and C$^{18}$O (109.78 GHz) in three narrowband spectral windows with a spectral resolution of 98 kHz, corresponding to $\sim$ 0.27 km s$^{-1}$.  After applying the quality assurance gain tables supplied with the data and line length corrections in \textsc{miriad} \citep{Sault11}, the data were imported into \textsc{casa} \citep{McMullin07}, and all further $uv$ plane data reduction was done in that program. Bandpass and gain calibration were done for each wideband spectral window, the gains of which were applied to the narrowband science data.  A further  bandpass calibration was then performed on the narrowband data, and applied along with flux and gain calibration from the wideband data. When Neptune was used as the flux calibrator, the `Butler-JPL-Horizons 2012' flux scaling standard was used.  When MWC349 was the flux calibrator, a 100 GHz flux of 1.3 Jy was assumed.  The total integration time, including time on calibrators, was $\sim$ 29 hours, 17 of which were on source.  
The seven datasets were then catenated together for combination with the FCRAO data described below. The resultant spatial resolution of these observations is 5.29$''\times4.48''$ at a position angle of 33.5 deg.  The largest angular scales probed by the CARMA data alone was approximately 60$''$.  The presence of larger structures in the observed region necessitated  including single dish data to recover all of the flux.

\begin{table}
\begin{center}
\caption{Calibrators used for each CARMA dataset.}
\begin{tabular}{rccccccc}
\hline
\hline
dataset:& 1&3&4&5&6&8&9\\
\hline
Flux:& N&  M & M&N & N  & M & M \\
Bandpass:&A &A &A&B & A  &A&B\\ 
Gain:&\multicolumn{7}{c}{2025+337}\\
\hline
\hline
\multicolumn{8}{l}{ The single digit codes above refer to: N - Neptune, }\\
\multicolumn{8}{l}{ M - MWC349, A - 2015+372, B - 3C454.3, C - 3C273}
\end{tabular}
\label{tab:calibrators}
\end{center}
\end{table}

\begin{figure}
\includegraphics[width=\columnwidth]{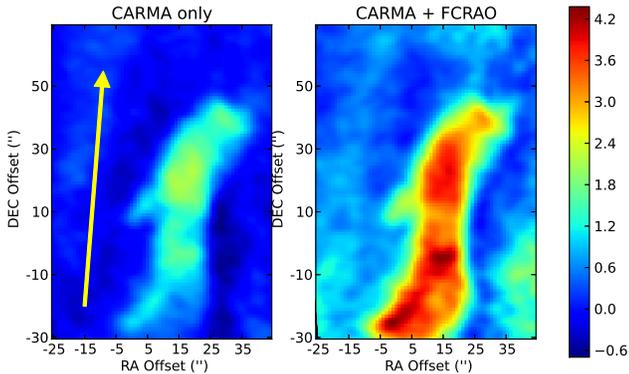}
\caption{$^{13}$CO integrated intensity maps for CARMA only (a) and from the combination of CARMA and FCRAO data (b).  The two maps are plotted on the same colour scale presented on the right. {\rev The yellow arrow shows the direction to the ionizing source.}}
\label{fig:combo_comparison}
\end{figure}

\subsection{FCRAO data}
\label{sec:FCRAO}

The FCRAO data used here are from the Exeter-FCRAO CO Galactic Plane Survey (Brunt et al., 2014, in prep.). The survey consists of fully sampled maps of $^{12}$CO and $^{13}$CO ($J$=1$-$0) with a spatial resolution of 45$^{\prime\prime}$ and 46$^{\prime\prime}$ respectively and a spectral resolution of ~0.15 km\,s$^{-1}$ over the regions 56$^{\circ}\leq~\ell~\leq$102$^{\circ}$, $\mid$b$\mid\gtrsim$1$^{\circ}$ and 142$^{\circ}\leq~\ell~\leq$193$^{\circ}$, -3.5$^{\circ}\leq~$b$~\leq$5.5$^{\circ}$. In the region 56$^{\circ}\leq~\ell~\leq$65$^{\circ}$, $\mid$b$\mid\gtrsim$1$^{\circ}$, the survey also contains C$^{18}$O ($J$=1$-$0) observations at the same resolution as the $^{13}$CO observations, which we also make use of here. The survey data have been corrected for stray-radiation, resulting in an absolute flux calibration accuracy of $\lesssim$10$\%$ (Mottram et al., 2014, in prep.). The median $\sigma_{rms}$ values for these data in the region studied in this paper are 1.5K, 0.4K and 0.4K in 0.127 km\,s$^{-1}$ channels for $^{12}$CO, $^{13}$CO and C$^{18}$O respectively.  The pixel size in these maps is 22.5$^{\prime\prime}$.

\subsection{Data Combination}
\label{sec:datacombo}

Because the structures probed in these observations are larger than the largest angular size observable with CARMA alone in its D configuration (see Section \ref{sec:carma}), we require short-spacing information from a single dish telescope. These data came from the FCRAO (as described above in Section \ref{sec:FCRAO}).  Both $uv$ and image plane image combination techniques were used, and we found that the $uv$ plane combination techniques worked best with our data.

  The frequency axis of the FCRAO data was regridded to the spectral resolution of the CARMA data, and the \textsc{CASA} task simobserve was then used to Fourier Transform the FCRAO data into the $uv$ plane.  The CARMA and FCRAO visibilities were then exported to $uv$ plane FITS files, and imported into \textsc{miriad} to be inverted together.   The two datasets were then jointly inverted, and then cleaned using the \textsc{miriad} task mossdi, and finally restored to give a final clean image.  

Figure \ref{fig:combo_comparison} shows the flux density of the CARMA only data compared with the flux densities of the combined dataset for $^{13}$CO.  The fluxes of the $^{13}$CO show the improvement in the flux recovery when the two datasets are combined.

\section{Results}
\label{sec:results}

 Here we present the physical and dynamical properties of the pillar we observed with CARMA and the FCRAO.  We begin by  quantifying the column density and temperature from previous {\it Herschel} observations of this pillar. We then describe the emission characteristics of the CO emission, specifically deriving the optical depth, velocity gradients and velocity dispersions of the three observed isotopologues.  We continue by determining the gas mass in the region as well as the fragmentation within the observed portion of the pillar which we relate to the Jeans length.

\subsection{Dust Properties}
\label{sec:dust}

\begin{figure}
\includegraphics[width=\columnwidth]{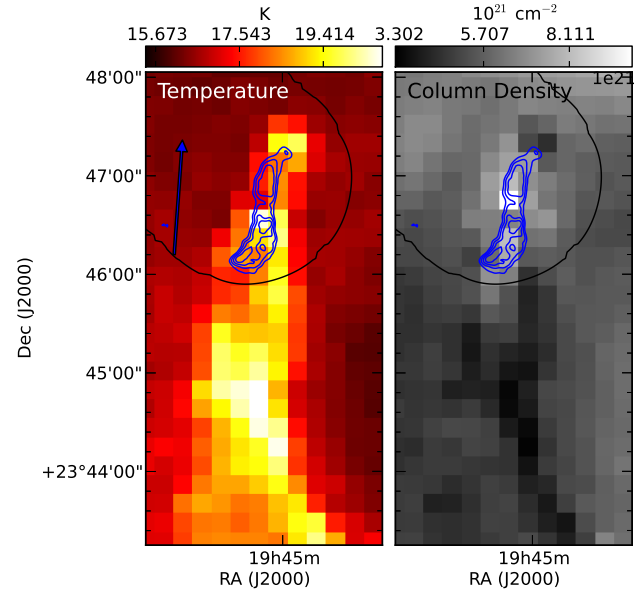}
\caption{Ambient temperature (left panel) and dust column density (right panel) derived from blackbody fits to the HiGAL data for the pillar. The blue contours correspond to the 60-90\% intensity levels of the \isot{13}CO emission we mapped in the pillar, with the single black contour showing the 60\% gain contour of the CARMA map. {\rev The blue arrow shows the direction to the ionizing source.}}
\label{fig:dust}
\end{figure}

To determine the dust column density and temperature structure in this pillar, we used a modified blackbody function to fit the { HiGAL} data for this region  \citep{Molinari2010}.  To facilitate this, we regridded the maps from the five bands (at 70, 160, 250, 350, and 500 $\mu$m) to the spatial grid  of the 500 $\mu$m data (11.5\arcsec pixels).  For each pixel, we then fit a modified blackbody function of the form

\begin{equation}
I_\nu =  \kappa_{\nu_0}(\nu/\nu_0)^\beta B_\nu(T)\Sigma
\end{equation}

where the dust opacity ($\kappa_{\nu_0}=0.1$ cm$^2 $ g$^{-1}$) is the dust opacity at a reference frequency ($\nu_0$ = 1000 GHz), the emissivity power law index ($\beta = 2$), and ($\Sigma = \mu $m$_h $N$($H$_2)$) are defined as in \citet{Sadavoy13}, and  $B_\nu(T)$ is the Planck function.  We fit the data using the Orthogonal Data Regression methods built into the \textsc{scipy} module of python. The resultant column densities and temperatures in the pillar are shown in Figure \ref{fig:dust}.  We find temperatures in the range of 17 to 22 K, with a mean of 18.2 K in the region covered by our CO maps.  We find column densities in the range of 4-10$\times10^{21}$ cm$^{-2}$, with lower column densities corresponding to regions of increased temperature.

From the column density map, we estimated the gas mass in the region observed with CARMA using:

\begin{equation}
M_{gas} = \left<N\right>*A*2.8*m_p
\label{eqn:mass}
\end{equation}

where $ \left<N\right>$ is the average gas column density derived from the greybody fit to the {\it Herschel} data in each pixel in the \isot{13}CO emitting area ($A$), and 2.8*$m_p$ is the mean molecular weight (as $m_p$ is the mass of a proton).  From this equation we find a gas mass of 91 M$_\odot$.

We note that for the portion of the pillar observed with CARMA, the dust temperatures and column densities are consistent {\rev along the long axis} of the pillar, suggesting that there is no difference in the pillar irradiation between the sides closer and further from the ionizing source.  This may be due to the low resolution of our temperature and column density maps, which are limited by the resolution of the 500$\mu$m Herschel dust map of the region (36.5$''$).

\label{sec:protostar}

From the 70$\mu$m emission map of this region, there appears to be an embedded core within our pillar.  It does not appear in any of the shorter wavelength observations of MIPSGAL or GLIMPSE \citep{Churchwell09} suggesting it is a cold, and embedded core.  Its position is shown with a black contour in Figure  \ref{fig:CO_moments}.  

\subsection{Line Emission Properties}
\label{subsec:emis}

\label{sec:layers}
 
\begin{figure*}
\includegraphics[width=\textwidth]{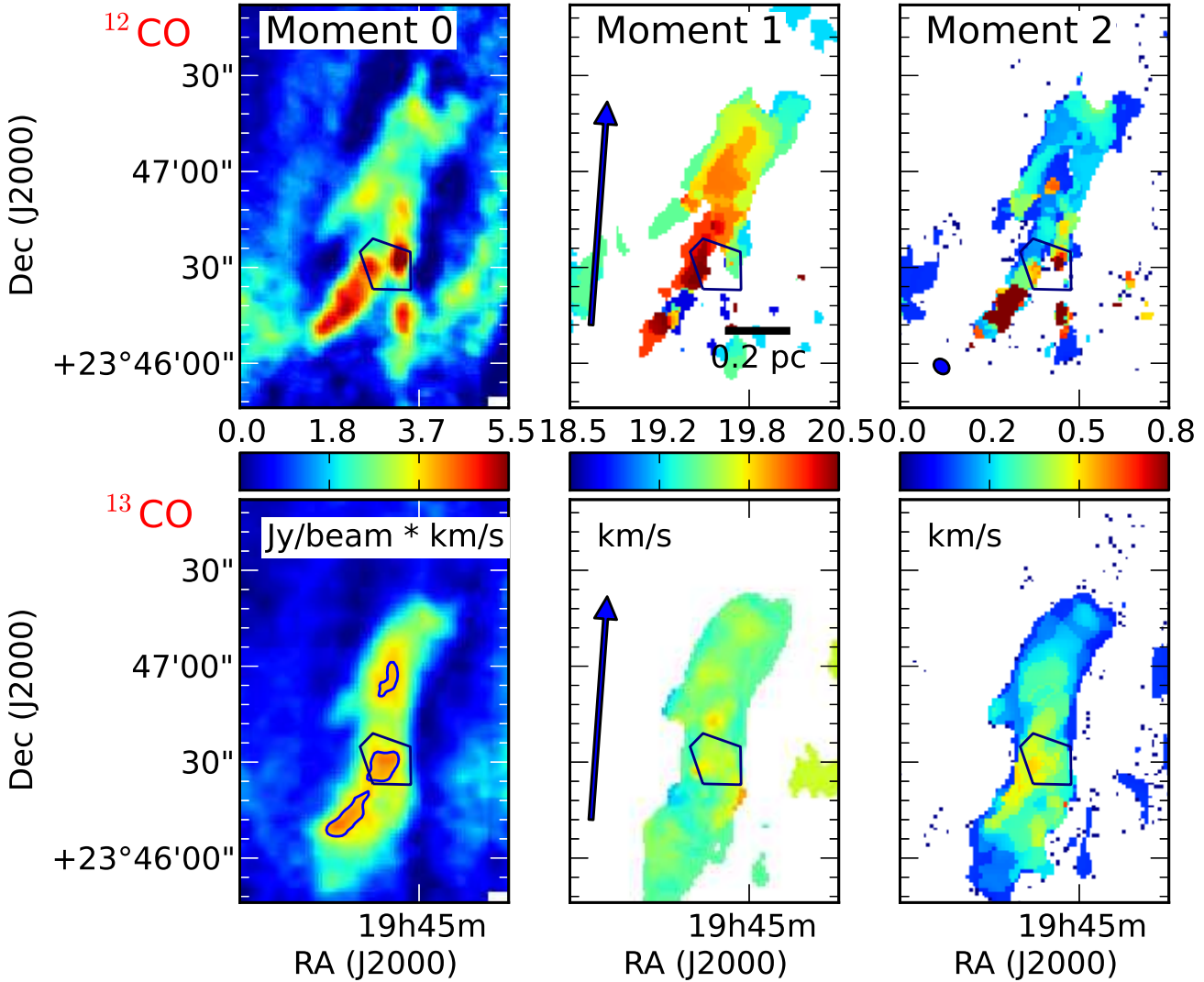}
\caption{The panels plotted here show (from left to right), the integrated intensity, the intensity weighted velocity and the velocity dispersion.The upper panels show these maps for \isot{12}CO, while the lower panels show them for \isot{13}CO. A scale bar is shown in the top middle panel, while the synthesized beam for these observations is shown in the top right panel. The colour scales are the same for the \isot{12}CO and \isot{13}CO maps, with the units of the scale given below each colorbar. The single black contour corresponds to the HiGAL 70$\mu$m emission of the embedded core in the pillar. The blue contours in the bottom left panel highlight the peaks in the \isot{13}CO emission and are drawn at 87\% of the peak intensity. {\rev The blue arrows in the middle panels show the direction to the ionizing source.}}
\label{fig:CO_moments}
\end{figure*}

 The brightest \isot{12}CO (shown in the top row of Figure \ref{fig:CO_moments}) appears at the edges of the pillar, with the \isot{13}CO (shown in the bottom row of Figure \ref{fig:CO_moments}) and C\isot{18}O emitting closer to the centre of the pillar. This is likely because the \isot{12}CO becomes optically thick quicker, and we are seeing limb brightened emission towards the pillar edges.  

{\rev
This kind of limb-brightening can be caused by temperature and/or linewidth gradients towards the center of the pillar. Since the outer layers are exposed to a stronger radiation field than the central regions of the pillar a negative temperature gradient is expected towards the center, which can cause the observed limb-brightening. The turbulence can also be stronger in the outer layers than
in the center of the pillar, which causes a linewidth gradient across the pillar. Unfortunately our data do not allow us to distinguish between these two mechanisms. To detect temperature gradients in/across the pillars we would need multiple transitions of the same molecule while the detection of linewidth gradient would require higher spectral resolution. 
}

\begin{figure*}
\includegraphics[width=\textwidth]{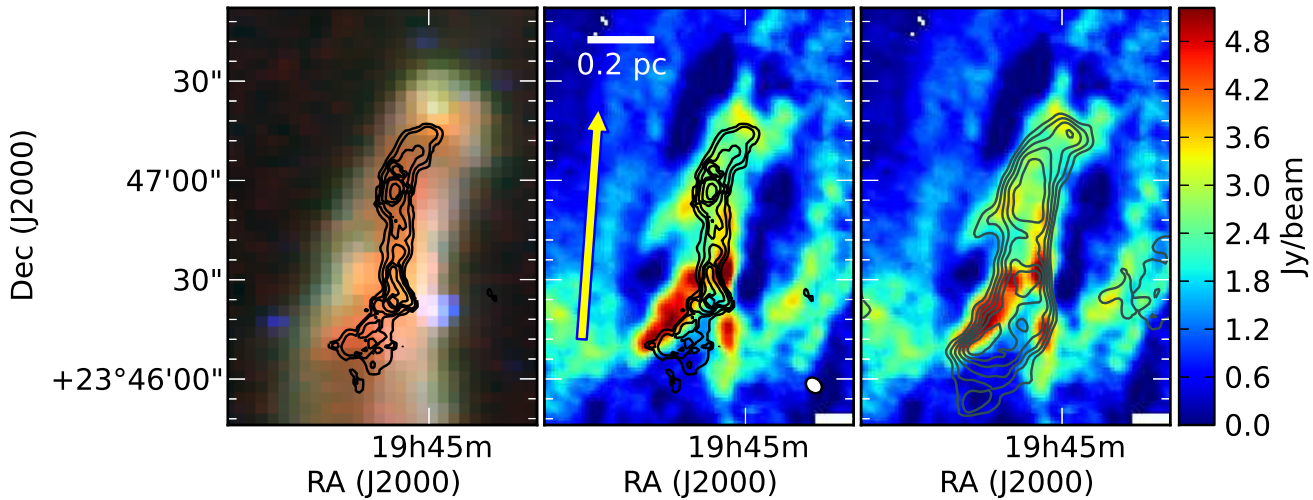}
\caption{The layered nature of the CO emission is seen here with the colour scale in the middle and right panels showing the integrated intensity of \isot{12}CO over plotted with the \isot{13}CO (right) and C\isot{18}O (centre) intensity starting at 40\% of the peak and increasing in 10\% intervals.  The three colour image on the left (also over plotted with the C\isot{18}O contours) shows the 8$\mu$m, 24$\mu$m and 70$\mu$m emission from the pillar tip. {\rev The yellow arrow in the middle panel shows the direction to the ionizing source.}}
\label{fig:layers}
\end{figure*}

\begin{figure}
\includegraphics[width=\columnwidth]{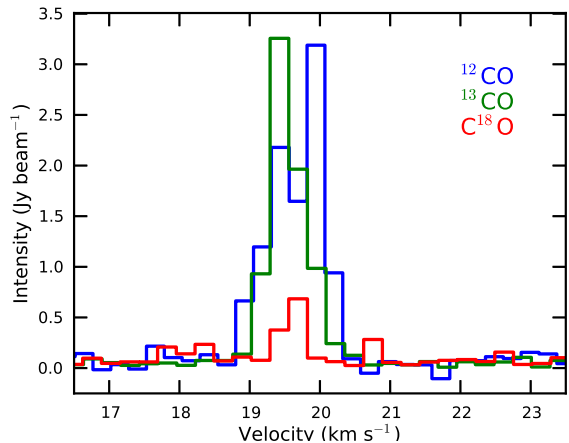}
\caption{CO and isotopologue spectra taken near the tip of the pillar. The \isot{12}CO has a double peaked profile, while the rarer \isot{13}CO and C\isot{18}O are singly peaked. This shows that the \isot{12}CO is likely optically thick.}
\label{fig:layer_spectrum}
\end{figure}

 The layered structure of the three isotopologues can be seen in Figure \ref{fig:layers}, with the \isot{13}CO emitting in the bulk of the pillar, and the C\isot{18}O showing the densest regions within it. In the middle panel of Figure \ref{fig:layers} the \isot{12}CO emission (colour scale) appears to peak directly exterior to the emission from C\isot{18}O (contours), suggesting that we are seeing layers of CO emission.  In Figure \ref{fig:layer_spectrum} we show a spectrum of the three CO isotopologues taken near the tip of the pillar. Assuming that the layered nature of the emission comes from optical depth effects, we calculated the optical depth at each point in the pillar.   This was accomplished using the \isot{13}CO and C\isot{18}O, as they trace the central mass of the pillar, whereas the \isot{12}CO emission becomes optically thick quite close to the edges of the pillar.  Assuming the C\isot{18}O was optically thin, we derived the \isot{13}CO optical depth from the relative strengths of the two lines at each velocity using a constant abundance ratio  of [\isot{13}CO]/[C\isot{18}O] = 7.7. At the rest velocity of the source ($\sim$19.8 km s$^{-1}$), the average optical depth of the \isot{13}CO line across the imaged region is $\tau\sim2$.

\label{sec:velgrad}
From the \isot{12}CO first moment map in Figure \ref{fig:CO_moments}, it appears as though there is a velocity gradient along the length of the pillar, with bluer material at the tip of the pillar, and redder material towards the base.  The \isot{13}CO first moment map (bottom middle of Figure \ref{fig:CO_moments}) does not show this velocity gradient. This suggests that the outer edges of the pillar (as traced by the \isot{12}CO) are being pushed away from the ionizing source, but that the core of the pillar (as traced by the \isot{13}CO and C\isot{18}O) is not.  That the less abundant isotopologues have minimal velocity gradients and low velocity dispersions suggests that this deeper gas has not yet been affected by the ionizing radiation.

{\rev
Across the pillar, perpendicular to the direction to the ionizing source, it appears that the \isot{12}CO is more blue shifted at the edges of the pillar than towards the centre. This is likely the effect of the photoionization of the pillar itself. This effect was also seen in a pillar near IC 1805 by \citet{Taylor99}, who were able to fit the velocity gradient by a simple shock model.

}

\label{sec:veldisp}
The velocity dispersion of the rarer isotopologues (\isot{13}CO and C\isot{18}O) appear to be roughly the same at 0.3-0.5 km s$^{-1}$ (close to our velocity resolution), with higher velocity dispersions seen towards the centre of the pillar (see the bottom right panel of Figure \ref{fig:CO_moments}).  The high ($\sim$ 1.4 km s$^{-1}$) velocity dispersion seen in the \isot{12}CO map either suggests truly higher velocity dispersions on the edges of the pillar, or that the \isot{12}CO is optically thick and the line shape is not truly indicative of the total gas motions.  Note that none of the features in the velocity dispersion map of the \isot{12}CO map correspond to any of the structures (knots) seen in the C\isot{18}O integrated intensity.  Indeed, the velocity dispersions seen in the \isot{13}CO emission appear to be anti-correlated with the C\isot{18}O features as shown with the contours overlaid on the velocity dispersion maps.  The C\isot{18}O is tracing the denser, and more quiescent  cores.

\subsection{Molecular Gas Properties}

 
From both the \isot{13}CO emission as well as the dust emission (see above), we find average column densities of approximately 8$\times10^{21}$ cm$^{-2}$ within the pillar.  Using the optical depth corrected \isot{13}CO emission map, and an {\rev excitation} temperature of 18 K (see Section \ref{sec:dust}), we used a modified version of equations 10.30 and 10.41 from \citet{Tielens05} for \isot{13}CO to determine the column density of the gas mapped in \isot{13}CO, which we converted to a total gas mass assuming a \isot{13}CO abundance of 1.4$\times10^{-6}$ {\rev\citep{Wilson94}} with respect to H$_2$, and integrating over the signal region in our map.  We find a gas mass of 94 M$_\odot$ from the region enclosed by the 3$\sigma$ contour of the \isot{13}CO integrated intensity.  This is in good agreement with the 91 M$_\odot$ found in Section \ref{sec:dust} from the dust.

 Two likely cores are seen in the portion of Vulp3  which we have mapped, which correspond to the northern two regions highlighted with contours in the \isot{13}CO integrated intensity map in the bottom left panel of Figure \ref{fig:CO_moments}.  One of these cores  corresponds to a HiGAL source at 70$\mu$m, as highlighted with a black contour in the same figure.  From the integrated intensities of the C\isot{18}O emission in these two specific regions, we calculated the column density of total gas assuming {\rev an excitation} temperature of 18 K, a C\isot{18}O abundance of 1.8$\times10^{-7}$ {\rev\citep{Wilson94}} with respect to H$_2$. The temperature comes from assuming the gas and dust are thermally coupled, and the abundance assumes that the CO is not frozen out at these temperatures. We then calculated the gas mass using equation \ref{eqn:mass}, and found that the more northern core has a mass of 14$\pm3$ M$_\odot$, and the more southern core (which corresponds to the 70$\mu$m source) has a mass of 16$\pm3$ M$_\odot$.

\section{Discussion}
\label{sec:discussion}

Below we discuss the implications of our observations and derived parameters, both in terms of the physics within the pillar, as well as in comparison to the predictions from simulations which produce pillars.

\subsection{Physical Processes}

\label{subsec:collapse}

In terms of the ongoing collapse of the pillar, \citet{Tremblin13} suggest that young pillars should show a double peaked spectral line profile indicating the collapse of the pillar inwards.  These two velocity peaks should be separated by twice the speed of the ionizing wind (corrected for the angle with respect to the line of sight), as each side of the pillar is undergoing collapse.  Barring the double peak seen in our \isot{12}CO emission, which is likely an optical depth effect, we only see single peaked profiles in the \isot{13}CO and C\isot{18}O emission in the observed pillar (See Figure \ref{fig:layer_spectrum}). This suggests that our pillar is not collapsing in on itself, and may be older than the pillars probed in their work.


From the column density in the pillar, and the assumption that the pillar is cylindrical, we derive a gas volume density of 8$\times10^{3}$ cm$^{-3}$. This comes from the gas column density derived from both the gas and dust, which are consistent with each other.  Further using the average temperature of 18.2 K within the pillar (see Section \ref{sec:dust}), we find a Jeans length of 0.3 pc ($\sim$ 27$''$ at a distance of 2.3 kpc) in the pillar.  This length is consistent with the distances between the peaks in the \isot{13}CO emission (highlighted with blue contours in Figure \ref{fig:CO_moments}). 
Interestingly, the small peak seen at the {\it very} tip of the pillar in the bottom left panel of Figure \ref{fig:CO_moments} is 0.22 pc from the next core (the topmost core highlighted by a blue contour).  Our Jeans analysis relies on the average temperature and density derived from the Herschel dust maps which have a resolution of $\sim$ 36.5\arcsec and are too coarse to determine whether there is a density enhancement at the tip of the pillar.  That the distance between the intensity peaks decreases to 0.22 pc ($\sim 20$\arcsec), near the tip suggests that the density is enhanced at the tip of the pillar, the likely result of the tip being compressed and photoionized.


The part of the pillar we mapped appears to have a mass of 94 M$_\odot$, and two embedded cores of approximately 15 M$_\odot$ each.  Assuming an upper limit for the star formation efficiency of 30\% within each core \citep[i.e.][]{Frank14}, we would expect this pillar tip to produce a few stars of $\sim1$ M$_\odot$ each.  This gives an overall star forming efficiency of less than, but not inconsistent with 10\% in the pillar tip. Thus, it does not seem likely that any triggering of star formation which takes place in this pillar will lead to a higher star formation efficiency than in molecular clouds, though it might speed up the processes through radiatively driven implosion.

\begin{figure}
\includegraphics[width=\columnwidth]{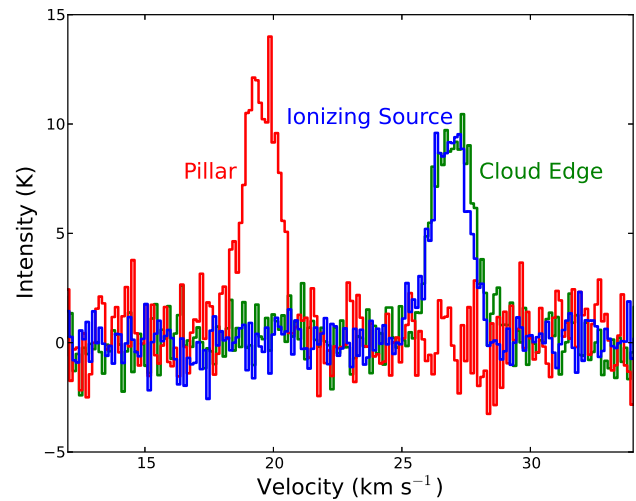}
\caption{Spectra of \isot{12}CO observed with the FCRAO taken towards single pointings in the region of the ionizing source, the portion of the pillar imaged with CARMA and the ambient cloud the pillar sticks out of. These spectra show that the pillar has a different systemic velocity than the cloud and ionizing source.}
\label{fig:Voffset_spectra}
\end{figure}

\begin{figure}
\includegraphics[width=\columnwidth]{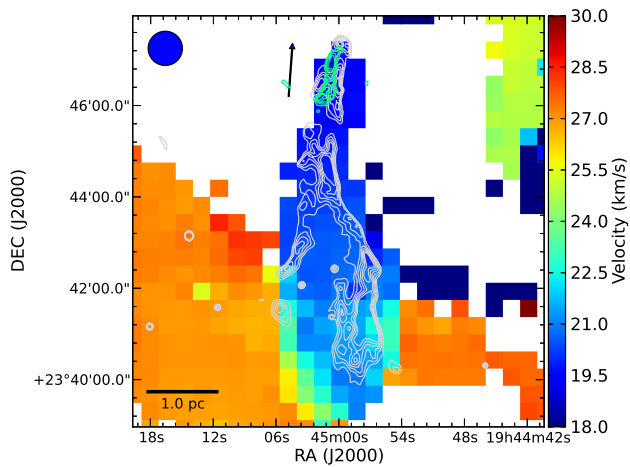}
\caption{Velocity structure (first moment map) of the ambient $^{12}$CO emission from the FCRAO map.   The grey contours show the MIPS 24$\mu$m emission structure, and the green contours show the region for which we have combined FCRAO and CARMA data (the contours correspond to the $^{13}$CO emission).  This shows that the pillar has a systemic velocity near 20 km s$^{-1}$, where as the cloud in which it is embedded appears to have systemic velocity closer to 26 km s$^{-1}$, suggesting the pillar is moving with respect to the cloud rim in which it is embedded. {\rev The black arrow shows the direction to the ionizing source, and is the same relative size as those shown in Figures 1-4.}}
\label{fig:ambient_FCRAO}
\end{figure}

The suggested ionizing source for Vulp3 is embedded in gas with a systemic velocity of $\sim$ 26 km s$^{-1}$.  This is the same velocity as the ambient gas near Vulp3, the gas at the edge of the cloud (see Figure \ref{fig:Voffset_spectra}).    In Figure \ref{fig:ambient_FCRAO} we show the large scale (FCRAO only) molecular gas motions of the pillar and nearby cloud edge. Vulp3 appears to be moving as a whole with respect to the cloud edge, and has an average velocity of $\sim$ 19-20 km s$^{-1}$. This can also be seen in the spectra of Figure \ref{fig:Voffset_spectra} which shows the \isot{12}CO spectra of the ionizing source, ambient cloud and pillar taken from the FCRAO data. 

{\rev
There are two possible explanations for the velocity shift between the pillar and the ambient material (both at the ionizing source and at the cloud edge).  The first is that the pillar is part of a separate cloud from the cloud edge which only appears to be connected with the cloud edge because of projection effects.  If the pillar were part of a separate cloud, this could explain the large velocity shift between it and the cloud edge.  The  second is that the pillar is being pushed away from the ionizing source into the ambient cloud.  We favour this second option because }both the cloud rim into which the pillar is being pushed, as well as the ionizing source have the same systemic velocity while the systemic velocity of the pillar is blue shifted by 6 km s$^{-1}$.

\subsection{Comparison with Model Predictions}

{\rev 
Here we directly compare our observations to the predictions of three theoretical models. To ensure an equitable comparison, we focus on three testable predictions common to each model: the internal velocity dispersion within the pillar, the velocity offset with respect to the ambient cloud, and internal (streaming) motions within the pillar itself. Each of the models have a unique combination of these parameters.  Below we give summaries of the three models and their testable predictions which we then compare with our observational results 


The models of \citet{Gritschneder09}  simulate plane-parallel radiatively driven implosion of a Bonner-Ebert sphere under assumption of `High', `Intermediate' and `Low' fluxes. The low-flux case forms a pillar-like structure after approximately 600kyr. The three testable predictions from these models are moderate (3-4 km s$^{-1}$) velocity dispersions, that the pillar is stationary with respect to the ionizing source and has no internal flows.  In our observed pillar, we see velocity dispersions much lower than those predicted, but do observe likely pillar motions with respect to the ambient cloud.  From our \isot{12}CO maps, we do see velocity structures within the gas, but these flows to not appear in the \isot{13}CO and C\isot{18}O moment maps suggesting they may only be surface features of the pillar.

  \citet{Gritschneder10} simulated the propagation of a planar ionizing flux into a box containing gas with turbulent velocity fields characterized by Mach numbers (relative to the sound speed in the cold gas) ranging from 1.5 to 12.5. They found that low Mach number turbulence was unable to produce any structures able to survive the propagation of the radiatively--driven shocks. However, they found that moderate to high Mach numbers generated a range of inhomogeneities which were rapidly eroded into pillar--like structures, several of which had dense gravitationally--unstable objects near their tips.  The three testable predictions from these models suggest velocity dispersions of order  1-2 km s$^{-1}$ with internal flows, and motion of the pillar itself with respect to the ambient cloud, being pushed away from the ionizing source at $\sim4-5$ km s$^{-1}$.

The velocity dispersions predicted in these models are consistent with those determined observationally.  We note that we may be seeing pillar motion, but it is unclear whether we observed internal flows of gas within the pillar.

  \citet{Dale12a,Dale12b,Dale13a} and \cite{Dale13b} modelled the propagation of ionizing radiation from O--stars from {\it inside} turbulent molecular clouds to examine the impact of feedback on the cloud structure and dynamics, and on the ongoing star formation process within the clouds. The simulated clouds have a filamentary structure generated by the interaction of turbulent flows. Gas flows along the filaments, so they are also accretion flows. The most massive clusters and stars therefore tend to form at filament junctions. Ionization feedback in these simulations often produces prominent bubbles on scales up to a few $\times10$pc. In several of the \citet{Dale12a} calculations, most prominently their Run I, prominent pillar structures emerge naturally from the erosion of the filaments/accretion flows by the massive stars. These pillars tend to have conical shapes.  The three testable predictions from \citet{Dale12a} are that there are velocity dispersions of order 1-2 km s$^{-1}$, internal flows within the pillar, but that the pillar itself is stationary with respect to the ambient material.
  
Our observed velocity dispersion is broadly consistent with those predicted by Dale et al. ($\sim$1 km s$^{-1}$), and our possible pillar motions are also consistent with their predictions. Judging from the \isot{13}CO first moment map, we are not seeing internal flows in the observed pillar, and Dale et al. do not see such motions in their simulations. However, there to appear to be velocity gradients along the long axis of the pillar in our \isot{12}CO emission. This is not seen in Dale et al.'s models, but is consistent with those of \cite{Gritschneder10}.  It seems that no theoretical model of pillar formation is yet able to reproduce all the observed characteristics, and that this process deserves more attention.

  

}

\section{Conclusions}
\label{sec:conclusions}

We presented high spatial resolution, molecular line observations of the tip of a pillar in Vulpecula being ionized by a nearby massive star.  We find the internal velocity dispersion of the gas to be quite low (0.3-0.5 km s$^{-1}$) near our spectral resolution of 0.3 km s$^{-1}$, and that the pillar itself is moving away from the ionizing source.  We also find that there may be internal motions within the tip of the pillar itself, which are unrelated to the protostellar cores observed in \isot{13}CO within our field of view.  

Our results are comparable to the testable predictions put forth in the models of \citet{Dale12a,Gritschneder10,Gritschneder09}, and we find that for two of these models \citep{Dale12a,Gritschneder09} a number of the predictions are consistent with our observations. However, the models of  \citet{Gritschneder10}, show only one consistent parameter with our observations, in that we see internal gas motions within the pillar.

\subsection*{Acknowledgements}

{\rev We would like to thank the referee, Mark Heyer, for his help in improving our manuscript. } Support for CARMA construction was derived from the states of California, Illinois, and Maryland, the James S. McDonnell Foundation, the Gordon and Betty Moore Foundation, the Kenneth T. and Eileen L. Norris Foundation, the University of Chicago, the Associates of the California Institute of Technology, and the National Science Foundation. Ongoing CARMA development and operations are supported by the National Science Foundation under a cooperative agreement, and by the CARMA partner universities. This research made use of APLpy, an open-source plotting package for Python hosted at http://aplpy.github.com. JCM is supported by grant 614.001.008 from the Netherlands Organisation for Scientific Research (NWO).

\bibliographystyle{mn2e}

\label{lastpage}

  \end{document}